\documentclass[showpacs,preprint,amssymb]{revtex4}
\usepackage{graphicx}
\usepackage{dcolumn}
\usepackage{bm}
\usepackage{amssymb}
\def\be{\begin{equation}} 
\def\ee{\end{equation}}
\def\bea{\begin{eqnarray}} 
\def\eea{\end{eqnarray}}
\def\line{\hbox to \hsize}    
\def\frac #1#2{{#1\over #2}}

\def\vev #1{{\langle #1\rangle}}
\def\1{\mbox{\bf 1}}

\begin{document}

\title{Wigner translations  and the observer-dependence of the position of masslesss spinning particles}

\author{ MICHAEL STONE}

\affiliation{University of Illinois, Department of Physics\\ 1110 W. Green St.\\
Urbana, IL 61801 USA\\E-mail: m-stone5@illinois.edu}   

\author{VATSAL DWIVEDI}

\affiliation{University of Illinois, Department of Physics\\ 1110 W. Green St.\\
Urbana, IL 61801 USA\\E-mail: vdwived2@illinois.edu}   

\author{TIANCI ZHOU}

\affiliation{University of Illinois, Department of Physics\\ 1110 W. Green St.\\
Urbana, IL 61801 USA\\E-mail: tzhou13@illinois.edu}

\begin{abstract}  The Wigner little group for massless particles is isomorphic to the Euclidean group ${\rm SE}(2)$.  Applied to momentum eigenstates, or to infinite plane waves, the Euclidean ``Wigner translations''  act as the identity.   We show that when applied to finite wavepackets the translation generators move the packet trajectory parallel to itself through a distance proportional to the particle's helicity.  We relate this effect to the  Hall effect of light and to the Lorentz-frame dependence of the position of a massless spinning particle.

\end{abstract}

\pacs{03.30.+p,  03.50.De, 11.30.Cp}

\maketitle

\section{Introduction}
\label{SEC:introduction}

The Poincar{\'e}  group provides  the fundamental kinematic symmetry
of a relativistic particle. As  a non-compact group,  all its unitary representations are infinite dimensional,  but  in a  famous paper \cite{wigner}  Wigner  showed that the   physically interesting  representations can be induced from {finite-dimensional}  unitary  representations of a {\it little group\/}, which is the   subgroup of  homogeneous Lorentz transformations  that leaves some reference  four-momentum $p^\mu_0$  invariant. The  representation space of the little group is the Hilbert space for the particle's spin.   If the particle has positive  mass $m$,  we may take as reference  the four-momentum in the particle's  rest frame where $p^\mu_0=(m, {\bf 0})$. The  little group then consists of the  space  rotations ${\rm SO}(3)$. For a  massless particle  there is no rest frame  and  the reference momentum must   be a null vector $p^\mu_0 \to  (|{\bf p}_0|, {\bf p}_0)$.    The little group  now  consists of   space rotations ${\rm SO}(2)$ about the three-vector ${\bf p}_0$, together with  operations that are generated by infinitesimal Lorentz  boosts  in directions  perpendicular to ${\bf p}_0$ combined with compensating infinitesimal rotations. Remarkably  the combined  operations mutually commute, possess all the algebraic properties of Euclidean translations,  and the   resulting little group is   isomorphic to  the symmetry group ${\rm SE}(2)$ of the two-dimensional Euclidean plane.    What is  being moved  by these   translation operations?  The answer given by Wigner is that they  move   \underline{nothing}:  if the translation generators    had a physical effect, the little-group representation would be infinite dimensional and the particle being described  would have   ``continuous spin'' ---   a property  possessed by  no  known   particle.  Indeed the Wigner translations   have no effect when applied to   plane-wave solutions of the massless Dirac equation, and act as gauge transformations when applied to   the vector potentials of  plane-wave solutions of Maxwell's equations  \cite{han-kim-son}. Consequently they  act as the identity on the momentum eigenstates  created by the operator-valued  coefficients of the plane-wave modes,  thus ensuring that       the  spin of a   massless particle is entirely specified  by  a  finite-dimensional  representation of the  ${\rm SO(2)}$ helicity subgroup \cite{weinberg}.

 It is the purpose of this paper to show that, while they have no effect on {\it infinite\/}  plane waves,  when applied to {\it finite-size\/}  wave packets    of non-zero  helicity --- and in particular to circularly polarized Gaussian packets   --- the Wigner translations do have an effect:  they shift    the wave packet trajectory  parallel to  itself.  
 This shift is related to the  relativistic Hall effect of light \cite{onoda,horvathy-spin-optic,horvathy-fermat} and to the observer dependence of the location of massless particles \cite{bliokh-nori}. It gives rise to   the unusual Lorentz covariance properties  found \cite{son3,duval-horvathy-wigner} in the chiral kinetic theory approach to anomalous conservation laws  \cite{stephanov,stone-dwivedi,dwivedi-stone} and is also the source of the difficulty of obtaining  a conventionally covariant classical mechanics for a massless spinning particle in a gravitational field \cite{stone-dwivedi-zhou,duval-horvathy-chiral}.  
 
 In section \ref{SEC:poincare} we will provide a suggestive   algebraic argument for a sideways shift. In section  \ref{SEC:paraxial} we will  show that the shift actually occurs in finite-width beam solutions to Maxwell's equations. In section  \ref{SEC:discussion} we will  discuss and resolve a potential  paradox implied by the trajectory displacement.

\section{Poincar{\'e}  algebra   and massless particles}
\label{SEC:poincare}

 As an indication  that  Wigner translations can have a physical effect, we briefly review  a well-known   \cite{balachandran-atre,skagerstam} realization  of the Poincar{\'e}   algebra   for massless particles  of helicity $\lambda$    in terms of quantum mechanical  position and momentum operators. 
 We   start from the  familiar commutators
 \be
 [\hat x_i, \hat  p_j]=i\hbar\, \delta_{ij}, \quad [\hat p_i,\hat p_j]=0,
 \label{EQ:canonical}
 \ee
 and use the fact that the spin of a massless particle is slaved to its direction of motion to motivate the definition of the  angular momentum operator as
 \be
 J_k= \epsilon_{klm} \hat x_l\hat  p_m + \lambda \frac{\hat p_k}{|\hat {\bf p}|}.
 \label{EQ:qm-ang-mom}
 \ee
 This unconventional definition preserves the usual  commutation relation 
 \be
[J_k, \hat  p_l] = i\hbar\, \epsilon_{klm} \hat  p_m.
 \ee
 However, in  order to recover 
  \be
 [J_k, \hat  x_l]= i\hbar\, \epsilon_{klm} \hat  x_m,
 \ee
 and 
 \be
  [J_k,J_l]= i\hbar\, \epsilon_{klm} J_m,
  \ee
 we need to modify the commutator of the position-operator  components  to 
 \be
 [\hat x_k,\hat x_l]= -i \hbar \lambda\,\epsilon_{klm}\frac{\hat  p_m}{|\hat {\bf p}|^3}.
 \ee
 Accepting that the position-operator  components no longer commute, we can still  use $p^0\equiv |{\bf p}|$ to define a generator of Lorentz boosts in direction $k$ as 
 \be
 K_k={\textstyle \frac 12}(\hat  x_k|\hat  {\bf p}| +|\hat {\bf p}|\hat  x_k).
 \ee
 These  generators  satisfy   the remaining relations of the  Lorentz Lie algebra
 \bea
 {} [J_k, K_l]&=& i\hbar\,\epsilon_{klm} K_m,\nonumber\\
{} [K_k, K_l]&=& -i\hbar\,\epsilon_{klm} J_m,
\label{EQ:lorentz}
 \eea
 and   act   as expected on the momentum components:
 \bea
 [K_k,|\hat  {\bf p}|]&=&i\hbar\, \hat p_k,\nonumber\\
{} [K_k, \hat  p_l]&=&i\hbar\,\delta_{kl}|\hat {\bf p}|.
\label{EQ:p-boost}
 \eea
 We  have therefore constructed a representation of the Poincar{\'e} algebra on a quantum-mechanical  Hilbert space.
 
When we extend the   algebra to include the position  operators,  things become  more complicated.  We find (at $t=0$)
 \be
 [K_k,\hat  x_l]=-i\hbar\left\{{ \frac 12}\left(\hat  x_k\frac{ \hat  p_l}{|\hat  {\bf p}|} +\frac{ \hat  p_l}{|\hat  {\bf p}|} \hat x_k\right)
 +\lambda \epsilon_{klm} \frac{\hat  p_m}{|\hat  {\bf p}|^2}\right\}.
 \label{EQ:x-boost}
 \ee
 Neither term is immediately familiar. The expression  in parentheses arises  because the underlying Hamiltonian formalism automatically  maintains  the non-Lorentz invariant condition $x_0=t$ \cite{hansson-regge-teitelboim}. The term containing the helicity $\lambda$  will be  more interesting.
 
We  select a  reference four-momentum $p^\mu_0=(|{\bf p}_0|,{\bf p}_0)$ where ${\bf p}_0=(0,0,p)$ and   obtain  the corresponding Wigner translation generators   as the boosts and compensating rotations given by 
 \bea
 \Pi_1&=&K_1 +J_2,\nonumber\\
 \Pi_2&=&K_2-J_1.
 \label{EQ:wigner}
 \eea
 From  (\ref{EQ:lorentz}) we see  that these generators   obey the ${\rm SE}(2)$ Lie algebra
 \be
 [\Pi_1,\Pi_2]=0, \quad [J_3, \Pi_1]=i\hbar \,\Pi_2, \quad [J_3,\Pi_2]=-i\hbar\,\Pi_1.
 \ee
From (\ref{EQ:p-boost}) and (\ref{EQ:x-boost}) we also see  that  $\hat  x_1$,  $\hat  x_2$,  and the ${\rm SE}(2)$ generators  collectively  leave  invariant the eigenspace with  eigenvalues $ {\bf p}= (0,0,p)$ and any fixed  $x_3$. Acting within  the particular invariant subspace  with $x_3=0$, we find that
 \be
 [ \Pi_k, \hat  x_l]= - i\hbar\, \epsilon_{kl3}  \frac{\lambda}{p}, \quad (k,l=1,2).
 \label{EQ:wigner-subspace}
 \ee
 In (\ref{EQ:wigner-subspace}) the  Wigner ``translations''   seemingly   effect  a  genuine   infinitesimal translation of  the $ x_1$, $  x_2$ coordinates in the $x_3=0$ plane, and hence a translation of the particle trajectory ${\bf x}(t)=(x_1,x_2, t)$ parallel to itself.  Is  this apparent displacement merely   an artifact of an unconventional representation of the Poincar{\'e} algebra, or does  it have  something to do with physics?
  
 In the next section we will use solutions of  Maxwell's  equations to  illustrate  that  this sideways shift is not just a mathematical curiosity, but corresponds to what occurs in nature --- the trajectory  of  a circularly polarized photon  is  observer-dependent and is translated parallel to itself by   an infinitesimal  Lorentz boost and aberration-compensating rotation.

\section{Paraxial Maxwell beams}
\label{SEC:paraxial}

We wish to consider the action of boosts and rotations on a finite-size photon wavepacket.  It will serve to  consider their effect on  finite-width laser beam in the paraxial approximation. We will  use units in which  $\mu_0=\epsilon_0=c=1$.

The scalar paraxial wave equation 
\be
\frac{\partial^2 \chi}{\partial x^2}+ \frac{\partial^2 \chi}{\partial y^2}+2ki \frac {\partial \chi}{\partial z}=0
\label{EQ:paraxial-equation}
\ee
is obtained from the full scalar wave equation 
\be
\frac{\partial^2 \phi}{\partial x^2}+ \frac{\partial^2 \phi}{\partial y^2}+\frac{\partial^2 \phi}{\partial z^2}- \frac{\partial^2 \phi}{\partial t^2}=0
\ee
by writing 
\be
\phi({\bf r},t)= \chi({\bf r}) e^{ik(z-t)}
\ee
and assuming that $\chi({\bf r})$ is sufficiently slowly varying that we can ignore its  second derivative $\partial^2 \chi/\partial z^2$ in comparison to the remaining terms in (\ref{EQ:paraxial-equation}).

The simplest solution of  eq. (\ref{EQ:paraxial-equation})   is   the Gaussian-beam  \cite{garg-book}
\be
\chi({\bf r})=\frac{1}{(z-iz_0)} \exp\left\{ - \frac{x^2+y^2}{2 w^2(z)} +i k  \frac{x^2+y^2}{2 R(z)}\right\},
\label{EQ:scalar-paraxial}
\ee
where
\be
w^2(z) =\frac{z^2+z_0^2}{k z_0};\quad R(z)= \frac{z^2+z_0^2}{z}.
\ee
In this solution the  beam is propagating in the $+z$ direction, the quantity $w(z)$ is the width of the beam at a  distance $z$ away from  its  waist,  and $R(z)$ is the radius of curvature of the wavefront passing through the point ${\bf r}=(0,0,z)$.  The width grows linearly with $z$ once $z\gg z_0$, and the angular half-width is $1/kw(0)$. The condition for the paraxial approximation to be accurate ($kz_0\gg1$) is  equivalent to the beam having small asymptotic divergence.  We will always be interested in the region $z<z_0$ where the beam is narrow and almost  parallel sided.

\begin{figure}
\includegraphics[width=7.0in]{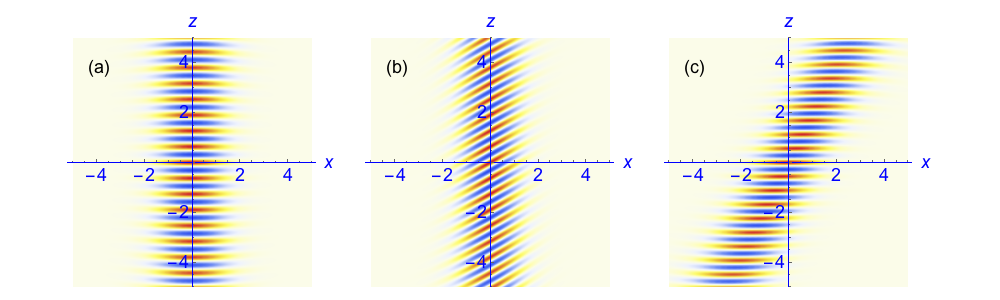}
\caption{\sl Slice through a  paraxial scalar beam with parameters $k=10$, $z_0=10$. {\rm  a)} Density plot of original beam amplitude  ${\rm Re}\{\chi(x,0,z,0)e^{ikz}\}$;  {\rm b)} Beam amplitude after Lorentz transformation (eq.\ (\ref{EQ:lorentz-transform}))  with rapidity $s=0.5$;  {\rm c)} Beam amplitude after both Lorentz transformation and aberration-compensating rotation though $\theta=- \tan^{-1}(\sinh s)=-31.5^\circ$. }
\label{FIG:fig1}
\end{figure}


From any two independent solutions $f$, $g$ of  the scalar paraxial  equation we can find \cite{erikson-singh} vector ${\bf E}$ and ${\bf B}$ fields that  are internally consistent  solutions of  Maxwell's equations  up to accuracy of order $1/(kl)^2$,   where $l$ is some charateristic length such as $z_0$ 
\bea
E_x({\bf r},t)&=&f({\bf r},t) +\frac 1{4k^2}\left(\frac{\partial^2 f}{\partial x^2}- \frac{\partial^2 f}{\partial y^2}\right)+\frac{1}{2k^2} \frac{\partial^2 g}{\partial x\partial y},\nonumber\\
E_y({\bf r},t)&=&g({\bf r},t)- \frac 1{4k^2}\left(\frac{\partial^2 g}{\partial x^2}- \frac{\partial^2 g}{\partial y^2}\right)+\frac{1}{2k^2} \frac{\partial^2 f}{\partial x\partial y},\nonumber\\
E_z({\bf r},t)&=&\frac{i}{k}\left(\frac{\partial f}{\partial x}+ \frac{\partial g}{\partial y}\right),
\label{EQ:paraxialE}
\eea
and
\bea
B_x({\bf r},t)&=&-g({\bf r},t) +\frac 1{4k^2}\left(\frac{\partial^2 g}{\partial x^2}- \frac{\partial^2 g}{\partial y^2}\right)+\frac{1}{2k^2} \frac{\partial^2 f}{\partial x\partial y},\nonumber\\
B_y({\bf r},t)&=&f({\bf r},t)- \frac 1{4k^2}\left(\frac{\partial^2 f}{\partial x^2}- \frac{\partial^2 f}{\partial y^2}\right)-\frac{1}{2k^2} \frac{\partial^2 g}{\partial x\partial y},\nonumber\\
B_z({\bf r},t)&=&-\frac{i}{k}\left(\frac{\partial g}{\partial x}- \frac{\partial f}{\partial y}\right).
\label{EQ:paraxialB}
\eea
To obtain a  Gaussian  ${\rm TEM}_{00}$ beam that is circularly polarized with  positive helicity  we take $f({\bf r},t)=\chi({\bf r}) e^{ik(z-t)}$ and $g({\bf r},t) =i\chi({\bf r}) e^{ik(z-t)}$, with $\chi({\bf r})$ given by eq.\ (\ref{EQ:scalar-paraxial}).

Using    Mathematica\textsuperscript{TM} to manipulate the resulting rather lengthy expressions we find, for example,  that the  time-average energy density in the beam is 
\be
T^{00}\equiv \frac 12 \vev{|{\bf E}|^2+|{\bf B}|^2}= \frac{(x^2+y^2 +4(z^2+z_0^2))^2}{8(z^2+z_0^2)^3}e^{-kz_0(x^2+y^2)/(z^2+z_0^2)},
\ee
and the three components of the time-averaged Poynting vector  ${\bf S}=\vev{{\bf E}\times {\bf B}}$ are 
\bea
S_x&=&T^{10}= \frac{(x^3z-x^2yz_0 +xy^2z -y^3z_0 +4(xz-yz_0)(z^2+z_0^2))}{2(z^2+z_0^2)^3}e^{-kz_0(x^2+y^2)/(z^2+z_0^2)},\nonumber\\
S_y&=&T^{20}=\frac{(y^3z +y^2xz_0+x^2yz +x^3z_0 +4(yz+xz_0)(z^2+z_0^2))}{2(z^2+z_0^2)^3}e^{-kz_0(x^2+y^2)/(z^2+z_0^2)},\nonumber\\
S_z&=& T^{30}=\frac{(-x^4 -2x^2y^2-y^4 +16(z^2+z_0^2)^2 )}{8(z^2+z_0^2)^3}e^{-kz_0(x^2+y^2)/(z^2+z_0^2)}.
\eea
The energy-flux streamlines  twist  in the direction of the beam  helicity \cite{berry-mcdonald}, consequently    the $z$ component of the angular momentum density 
\be
\Sigma_z=xS_y-yS_x
\ee
is non-zero.   If we integrate over the plane  $z=0$ we find that 
\be
P_z\stackrel{\rm def}{=} \int\!\!\!\int_{z=0} S_z \,dxdy= \frac{\pi(-1+8k^2 z_0^2)}{4k^3 z_0^3}= \frac{2}{k z_0}\left\{1+ O\left(\frac{1}{(kz_0)^2}\right)\right\},
\ee
and 
\be
J_z\stackrel{\rm def}{=}\int\!\!\!\int_{z=0} \Sigma_z \,dxdy=\frac{\pi(1+2kz_0)}{k^3 z_0^2}= \frac 1 k \frac{2}{k z_0}\left\{1+ O\left(\frac{1}{(kz_0)^2}\right)\right\}.
\ee
The ratio $P_z/J_z$  is equal to $k$  in region  ($kz_0\gg1$)   where   paraxial aproximation is accurate.  This is  what  is  to be expected: $P_z$  gives  the linear momentum per unit length, which should be  $\hbar k$ per photon; 
 $J_z$  gives  the  angular momentum per unit  length of the beam,  which should be  $\hbar$ per photon.
 
We now compute  the ${\bf E}$ and ${\bf B}$ fields  as seen from   a reference frame moving  along the $+x$ axis at rapidity $s$.  The corresponding Lorentz transformation takes  
\bea
E_x(x,y,z,t) &\mapsto& 
   E_x(x' ,y, z, t'),\nonumber\\
   E_y(x,y,z,t) &\mapsto &  
   E_y(x', y, z, t') \cosh s  - 
B_z(x', y, z, t',)     \sinh s\nonumber\\
   E_z (x,y,z,t) &\mapsto& 
  E_z(x', y, z, t')   \cosh s + 
    B_y(x', y, z, t')  \sinh s,\nonumber\\
 B_x(x,y,z,t) &\mapsto&
   B_x(x', y, z, t')\nonumber\\
B_y(x,y,z,t) &\mapsto&
B_y(x', y, z, t')   \cosh s + 
    E_z(x', y, z, t')  \sinh  s,\nonumber\\
  B_z(x,y,z,t) &\mapsto&
   B_z(x' y, z, t')  \cosh s- 
 E_y(x', y, z, t')   \sinh s,
   \label{EQ:lorentz-transform}
 \eea
where
\bea
x' &=& x\, \cosh  s+ t\, \sinh s\nonumber\\
t' &=& t\,\cosh s- x\, \sinh s.
\eea

The Lorentz transformation changes the wave vector   from ${\bf k}=(0,0,k)$ to ${\bf k}'=(k\, \sinh  s,0,k)$,  so the direction of propagation has been  rotated  though an aberration angle of $|\theta| =\tan^{-1}(\sinh s)$.  The wavefronts are therefore tilted. The beam   envelope, however,  still lies parallel to the $z$-axis, and  is moving towards the observer  at speed $\beta=\tanh s$ (see figure \ref{FIG:fig1}-b).

The Lorentz transformation also affects the energy density distribution and the Poynting-vector flux though the $z=0$  plane.  In addition to a Lorentz contraction it noticeably shifts the position of their maxima (see figure \ref{FIG:fig2}). To  quantify these shifts  we   can compute the location of the Lorentz transformed energy density and  energy flux centroids.
The required integrals are still Gaussian and can be done analytically. With  the definition
\be
{\mathcal E}=  \!\!\int\!\!\!\!\int_{z=0}  T^{00} dxdy,
\ee
we have 
\bea
[\Delta y]_{\rm density}&=& \frac{1}{\mathcal E}\int\!\!\!\!\int_{z=0} y \,T^{00} dxdy,\nonumber\\
&=&
\frac{z_0(4+8kz_0 \sinh s)}{(1+8kz_0+8k^2z_0^2) \cosh s -4kz_0 \,{\rm sech\,}s},\nonumber\\
&=&\frac 1 k \tanh s\left\{1+O\left(\frac{1}{(kz_0)^2}\right)\right\},
\eea
and 
\bea
[\Delta y]_{\rm flux}&=&\frac 1{P_z}\int\!\!\!\!\int_{z=0} y \,S_z dxdy,\nonumber\\ 
&=& 
\frac{2 z_0 (1 - 2 k z_0) \tanh s}{1 - 8 k^2 z_0^2},\nonumber\\
&=& \frac 1 {2k} \tanh s\left\{1+O\left(\frac{1}{(kz_0)^2}\right)\right\}.
\eea
For positive helicity, both centroids are displaced to the left when seen from an observer moving towards the upward-propagating beam. The   centroids do not coincide, the energy-flux centroid moving   only half as far as the energy-density centroid.   Such  displacements  are  not restricted   to Gaussian beams. A   similar boost-induced  sideways  shift  and centroid separation  was  exhibited  in \cite{bliokh-nori}  for  Bessel  beams possessing  orbital angular momentum.   It was  also explained there   that the centroid    separation  arises solely from the   geometrical effect  pointed out in \cite{aiello}:  because of their  corkscrew trajectories, energy-flux streamlines  passing through a surface   rotated away from  perpendicular   to the direction of propagation find themselves  inclined at different angles to the surface to  the right and left  of the plane of rotation. Consequently,  even in the absence of a Lorentz boost, the energy-flux centroid of a tilted beam is   displaced with respect to its energy-density centroid \cite{aiello}.  

We  wish to obtain a  finite-displacement version of the Wigner translations,  so, after performing the boost,  we rotate  the  Lorentz transformed   beam about ${\bf r}=0$ though an aberration-compensating angle of $\tan^{-1}(\sinh s)$. After the rotation  the wavevector becomes  ${\bf k}=(0,0,k\, \cosh s )$ and   the  wavefronts  again lie parallel to  the $x$-$y$ plane. Consequently   the energy-flux streamlines no longer possess  a left-right asymmetry. We find  numerically that the   position of the energy centroid in the $z=0$  plane  is unchanged by the rotation ($T^{00}$ is a scalar under space rotations) while  the energy-flux centroid moves into  coincidence  with the energy-density centroid.  
Thus,   as result of the combined boost and compensating rotation  {\it both\/}  centroids  have been shifted though  a distance $\Delta y= (1/k)\tanh s = \beta/k$, where $\beta=v/c$.  The beam spot is restored to its pre-boost   appearance, and  we could repeat the operation and translate the beam spot  through a further distance.  If we reverse the helicity, we change the sign of this shift.  

\begin{figure}
\includegraphics[width=5.0in]{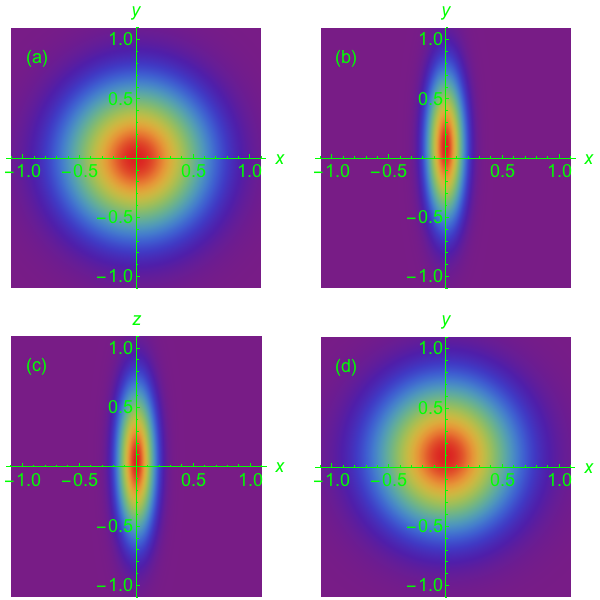}
\label{FIG:fig2}\caption{\sl Beam spot profiles   in the $z=0$  plane for  $k=10$, $z_0=3$.  {\rm  a)} Original intensity $ T^{00}(x,y)$;  {\rm b)} Lorentz transformation of $T^{00}(x,y)$ under eq.\ (\ref{EQ:lorentz-transform})  with rapidity $s=2.0$. The spot center is at $y=0.095$;  {\rm c)} Poynting energy flux $S_z(x,y)=T^{30}$  after Lorentz transformation. The spot center  is at $y=0.0475$; {\rm  d)} Poynting energy  flux $S_z(x,y)=T^{30}(x,y)$ after   aberration-compensating rotation. The flux maximum is at $y=0.095$.  The rotated intensity distribution has   similar appearance, and its  maximum is also at $y=0.095$.}
\label{FIG:fig2}
\end{figure}

In the absence of the lateral shift, the   combination of  boost and compensating rotation would leave  the trajectory of a short wavepacket  emitted from ${\bf r}=0$ at $t=0$ unchanged.  The continuous  beam, which can be though of  as arising from a stream of sequentially emitted wavepackets, is {\it  not\/}  left invariant, however.  How it changes is  shown in fig.\ \ref{FIG:fig1}-c. We see that the transformed beam  can  be though of as a sequence of pulses each fired in the $+z$ direction by an emitter that is moving rapidly to the left.  It is reminiscent  of a  diagonal steam of  strictly upward-moving  projectiles fired from a horizontally moving gun in the old Atari\textsuperscript{TM}   game ``Space Invaders.''    Any   particular packet continues to move parallel to the $z$ axis, but as a result of the lateral shift in the $z=0$ plane, its entire trajectory   is shifted sideways by 
$\Delta y=(1/k) \tanh s$. Figure \ \ref{FIG:fig1}-c also shows why the action of the Wigner translations take their simple form (\ref{EQ:wigner-subspace}) only  in the plane $x_3\equiv z=0$. In any other plane the translations get mixed up with the geometric effect of the rotation.  

The finite-$s$  boosts considered in this section have effects on the photon energy and intensity  that  appear  at quadratic order in the rapidity $s$. If we alternate a sequence of infinitesimal boosts and compensating rotations, the quadratic terms can be neglected and only the sideways shift (now equal  to  $\lambda/p$ times the total rapidity change)  remains.  We are in effect assembling a Trotter-product approximation that converges to  exponentials of the  Wigner translation generators    (\ref{EQ:wigner}).

\section{Discussion}
\label{SEC:discussion}

\begin{figure}
\includegraphics[width=2.2in]{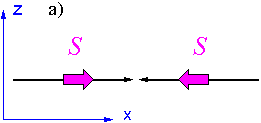}\includegraphics[width=2.2in]{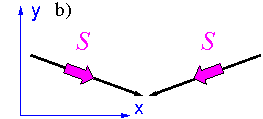}\includegraphics[width=2.2in]{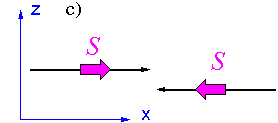}
\caption{\sl {\rm a)} A pair of  massless particles with spin ${\bf S}$ collide head-on;  {\rm b)} The particles viewed from above in  a frame moving towards the collision;  {\rm c)} A front view  from the moving frame shows the particles  miss one another.   }
\label{FIG:collision}
\end{figure}

The direction  and magnitude of the boost-induced lateral shift can be understood from a  geometric  picture  (See \cite{son3} for a related argument). Consider two massless particles, both possessing  helicity ${\bf p}\cdot {\bf S}_{\rm spin}/|{\bf p}|= \lambda$ and  heading directly  towards one another  parallel to  the $x$ axis. Because they will collide head-on, they have no relative orbital angular momentum and the two spin angular momenta ${\bf S}_{\rm spin}=( \pm  \lambda, 0 ,0)$ also sum to zero. Seen from a frame moving along the $y$ axis towards the collision point, however, the unit vectors in the  direction of the particles'   motion  have components   
$ (\pm {\rm sech\,} s ,-\tanh s ,0)$. Because the spin of a  massless  particle is slaved to its  direction of motion  there is now a net spin component of $2 \lambda  \tanh s$ directed towards the observer. Nonetheless, in  the new frame, the  total  angular momentum  will remain zero so   the spin contribution must be offset by an orbital angular momentum of $-2\lambda  \tanh s$. This orbital angular momentum can only come from a lateral shift of each particle's  trajectory by $\Delta z=(\pm \lambda/| {\bf p}|) \tanh s $ (see fig.\ \ref{FIG:collision}). For a photon ${\bf p}= \hbar\, {\bf k}$ and $\lambda=\hbar $,  so  we recover  the shift seen in our Gaussian beam.  Of course, if two particles   collide and produce two pions in one frame they must  produce two pions when seen from  another frame. That the particles  apparently miss  each other  because of the  sideways shift  cannot   affect  the pion production. The incipient paradox is resolved by the fact that   partial-wave scattering  amplitudes  depend  only   on the total relative angular momentum ${\bf J}={\bf L}+{\bf S}_{\rm spin}$ of the particles, and this quantity is not affected  by the shift.  The shift still has physical consequences, though.  If we move a detector  such as a photographic emulsion though the beam, it will be sensitive to either the energy density or the energy flux in its own rest frame, and these quantities  have  been displaced  by the motion. 

\section{Acknowledgements}   This work was supported by the National Science Foundation under grant number NSF DMR 13-06011. In the course of this work MS has exchanged many useful emails with Peter Horv\'athy and Christian Duval.   MS would also like to thank  Konstantin Bliokh for  discussions about the observer dependence of the location of light rays, and for drawing our attention to \cite{bliokh-nori}. We also thank the authors of \cite{son3} for sending us an early version  of their work.

\end{document}